\documentclass[aps,pra,reprint,twocolumn,superscriptaddress,,showpacs,showkeys,longbibliography]{revtex4-1}

\usepackage{amsmath,amssymb,amstext}
\usepackage[usenames,dvipsnames]{color}
\usepackage{graphicx}
\usepackage{bm,bbold,braket}
\usepackage{natbib}
\usepackage{txfonts, comment,stmaryrd}
\usepackage{dcolumn}

\usepackage[english]{babel}

\usepackage[colorlinks,bookmarks=false,citecolor=blue,linkcolor=red,urlcolor=blue]{hyperref}

\begin{document}
\title{A compact atomic beam based system for Doppler-free laser spectroscopy of Strontium atoms}
\author{Gunjan Verma}
\email {Author to whom correspondence should be addressed. Electronic mail: gunjan.verma@students.iiserpune.ac.in}
\affiliation{Department of Physics, Indian Institute of Science Education and Research, Pune 411008, Maharashtra, India}
\author{Chetan Vishwakarma}
\affiliation{Department of Physics, Indian Institute of Science Education and Research, Pune 411008, Maharashtra, India}
\author{C. V. Dharmadhikari} 
\affiliation{Department of Physics, Indian Institute of Science Education and Research, Pune 411008, Maharashtra, India}
\author{Umakant D. Rapol}
 
\affiliation{Department of Physics, Indian Institute of Science Education and Research, Pune 411008, Maharashtra, India}
\affiliation {Center for energy sciences, Indian Institute of Science Education and Research, Pune 411008, Maharashtra, India}


\begin{abstract}
We report the construction of a simple, light weight and compact atomic beam spectroscopy cell for Strontium atoms. The cell is built using  glass blowing technique and includes a simple Titanium Sublimation Pump for active pumping of the residual and background gases to maintain ultra-high vacuum. Commercially available and electrically heated dispenser source is used to generate the beam of Sr atoms. We perform spectroscopy on the  $5s^2\ ^1S_0\longrightarrow 5s\ 5p\ ^1P_1$ transition to obtained well resolved Doppler free spectroscopic signal for frequency stabilization of the laser source. This design can be easily extended for other alkali and alkaline earth metals.
\end{abstract}



\maketitle
The unprecedented fractional uncertainties in frequency measurements achieved in  optical clocks has led to efforts towards redefinition of the `second'. Strontium (Sr) atomic clock in particular is the foremost candidate in neutral atoms' based atomic clocks. Sr based clocks have reached fractional stability of 2.2 $\times$ 10$^{-16}$ and a total uncertainty of 2.1 $\times$ 10$^{-18}$ \cite{nichol2015}.  Efforts are also going on in building next generation space based optical clocks for fundamental physics, geophysics and astronomical applications \cite{schiller2012space, poli2014transportable}. For any such frequency reference to become a frequency standard, a large network of terrestrial and space based clocks is desirable leading to  many technical challenges. One of these challenges is to have a compact, stable and light weight system design. Lasers required for these  experiments, need to be stabilized to atomic spectral lines. However, unlike alkali atoms (Rb, K, Na), vapor pressure of Sr is very low at room temperature in addition to these atoms being highly reactive with glass\cite{bridge2009vapor}. This makes  spectroscopy of these atoms challenging. In earlier works on Sr spectroscopy, one of the following methods were used for frequency stabilization: a) An atomic beam derived from a high-flux Sr atomic oven, used to load atoms into a Magneto-optic trap was interrogated by an orthogonal laser beam to obtain the fluorescence spectrum \cite{schioppo2012compact} or b) Performing Doppler-free Saturated Absorption Spectroscopy (SAS) in a vapor cell \cite{bridge2009vapor}. There are several limitations of using these designs. One of the disadvantage of the system in \cite{schioppo2012compact} is that, the system gets more complex and bulky if one is only interested in performing independent laser spectroscopy. The other disadvantage is that, the atoms that are being interrogated are illuminated simultaneously by a Zeeman slowing laser beam, which leads to light-shift in the measured spectrum. In  ref. \cite{bridge2009vapor}, a more complex Doppler-free SAS was implemented to get a well resolved spectrum.

In this article, we report an improved design of a compact Sr spectroscopy cell  with respect to weight, size, complexity and stability of the previously reported system to derive well resolved Doppler-free spectrum of Sr atoms. Our system is completely built with glass and uses a commercial dispenser source and incorporates a compact low-power Titanium Sublimation pump (TSP) to maintain Ultra-High Vacuum (UHV). Our system is simpler to operate and in construction, which can be easily adapted to other elements of low vapor pressure elements by changing the dispenser source.

The schematic diagram of the cell is shown in Fig. 1, It consists of a main glass tube having an outer diameter (OD) of 19.2 mm and length 274 mm. An optical cross is created on the other end by fusing four glass tubes having an OD 28.5 mm and length 50 mm. Optical flats (windows) are fused on these tubes for optical access from orthogonal directions. Excitation of the atomic beam and collection of fluorescence from atoms is done from these two orthogonal directions. The cross eliminates the risk of forming dark coating by the deposition of Sr atoms on the glass windows that can block the optical access by probe beam.
We generate a well collimated atomic beam in the direction along the length of the long glass tube. We maintain a UHV ($< 10^{-8}$ mbar) in the entire glass cell by using a homemade TSP on the side of the long glass tube as shown in the Fig.1. This whole system is connected to a CF35 glass-to-metal adapter through a narrow neck for connecting it to pumping station based on Turbo-Molecular Pump (TMP).\par 
The filament for TSP is made by winding a 0.25 mm diameter Titanium wire onto a  Tungsten wire of 0.5 mm diameter and of length $\sim 90$ mm in the form of loop as shown in the inset of Fig. 1. The Tungsten loop provides physical support to Titanium wire and also ensures electrical conductivity within the Titanium wire. The electrical contacts for this filament is provided by another pair  of Tungsten wires of diameter 1 mm which is spot welded to the 0.5 mm wire. This TSP filament is enclosed in a spherical glass bulb of diameter $\sim$79 mm. The pumping efficiency of TSP is directly proportional to the internal surface area of this bulb.


\begin{figure}[hbt]
\centering
\includegraphics[width=0.7\linewidth]{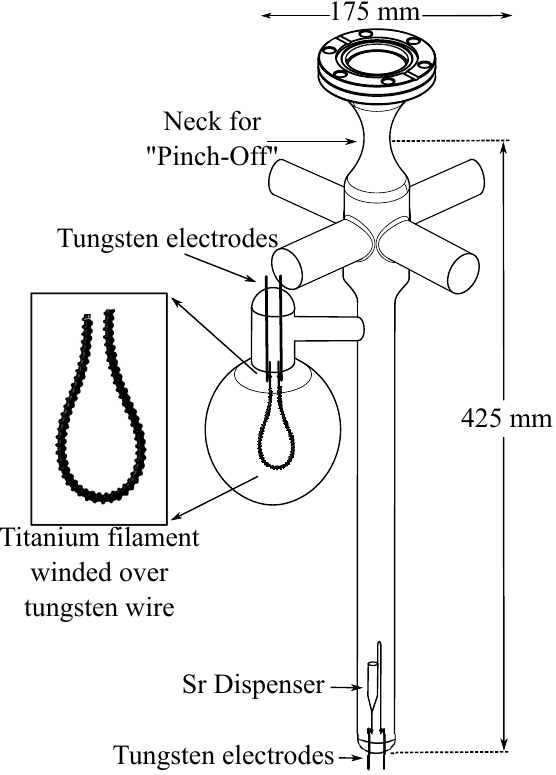}  
		\caption{Schematic diagram of spectroscopy cell.}
\label{fig:d1} 
\end{figure}
Tungsten electrodes cannot be fused directly to Pyrex glass as they have different coefficients of thermal expansion  ($4\times10^{-6}$m/m-K for glass and $4.3\times10^{-6}$m/m-K for tungsten) which can cause cracks and leaks in vacuum system. Hence, we use commercial Pyrex to Tungsten graded-glass in between the Tungsten and Pyrex glass and fuse the feedthroughs on the glass tube.\par
We use a commercial dispenser source form Alvatec (Alvasource AS-5-Sr-500-V) for generating the atomic beam of Sr atoms similar to the one reported in Ref. \cite{bridge2009vapor} albeit of a smaller length. This dispenser is a small capsule made up of stainless steel of OD 3 mm and length 25 mm. The dispenser is gas-tight sealed by pressing an Indium plug, and filled with isotopic mixture of Sr under pure argon atmosphere. When used, the source is heated via conventional resistance heating. As a consequence, the indium sealing melts and the small argon puff of gas is released in the vacuum system. During operation, the dispenser releases 99.95\% pure Sr  when the activation current is reached. The electrical connection to the dispenser is provided through a pair tungsten wires, which are spot welded with the legs of dispenser and fused with Pyrex glass.\par 

To achieve UHV, the spectroscopy cell is initially connected to the TMP pumping station using CF35 flange. An all-metal in-line valve is placed between the spectroscopy cell and the pumping station. The system is initially soft baked at a temperature of 140$^{\circ}$C for 24 hours while running the TMP continuously. The baking temperature is limited by the melting point of indium seal of the dispenser. During the bakeout cycle, the TSP filament and dispenser source are degassed by passing a current of 3 A and 2 A respectively. Before detaching the cell from the pumping station, the dispenser source is activated by melting the indium seal. This is done by gradually increasing the current through the dispenser from 3.5 A to 8 A, during this process we observe increase in pressure of up to $1.2\times10^{-3}$ mbar and then quick drops to its original value ($1.0\times10^{-5}$), caused by the  release of Argon from dispenser. The entire glass cell is gradually cooled down to room temperature to get the pressure below  $10^{-8}$ mbar. A Helium leak test was carried out for vacuum integrity at this point. The TSP filament is then flashed at a current of 4.5 A for a period of nearly 1 hour. This leads to improvement of vacuum by an order of magnitude. The spectroscopy cell is detached from TMP by closing the in-line valve  and slowly heating the ``neck'' of the cell with a gas torch and then ``pinching-off'' at that point.\par 
To characterize the spectroscopy cell, we excite the $5s^2\ ^1S_0\longrightarrow 5s\ 5p\ ^1P_1$ transition (saturation intensity, $I_s = 42.7 $mW/$cm^2$) of Sr atoms using a linearly polarized probe beam of 0.82 mW with a $1/e^2$ diameter of $\phi = 2$ mm at wavelength 460.7 nm (saturation parameter $=I/I_s = 0.61$). Atomic beam is generated by passing a current of 8.0 A through the dispenser. Due to finite dimensions of dispenser capsule, we get the intrinsic divergence of 120 mrad, this divergence is further reduced by the long glass tube of the cell to 95 mrad. We measure the longitudinal velocity of the atomic beam by having another probe beam at an angle of (4.7$^{\circ}$) and extract the most probable velocity to be 420 m/s. Fig. 2 shows the dependency of longitudinal velocity on source current. The probe light was generated from a home built frequency doubler which was fed with an input light of 922 nm with 1.2 W power generated from a commercial Tapered amplifier.
\begin{figure}[hbt]
\centering
\includegraphics[width=0.75\linewidth]{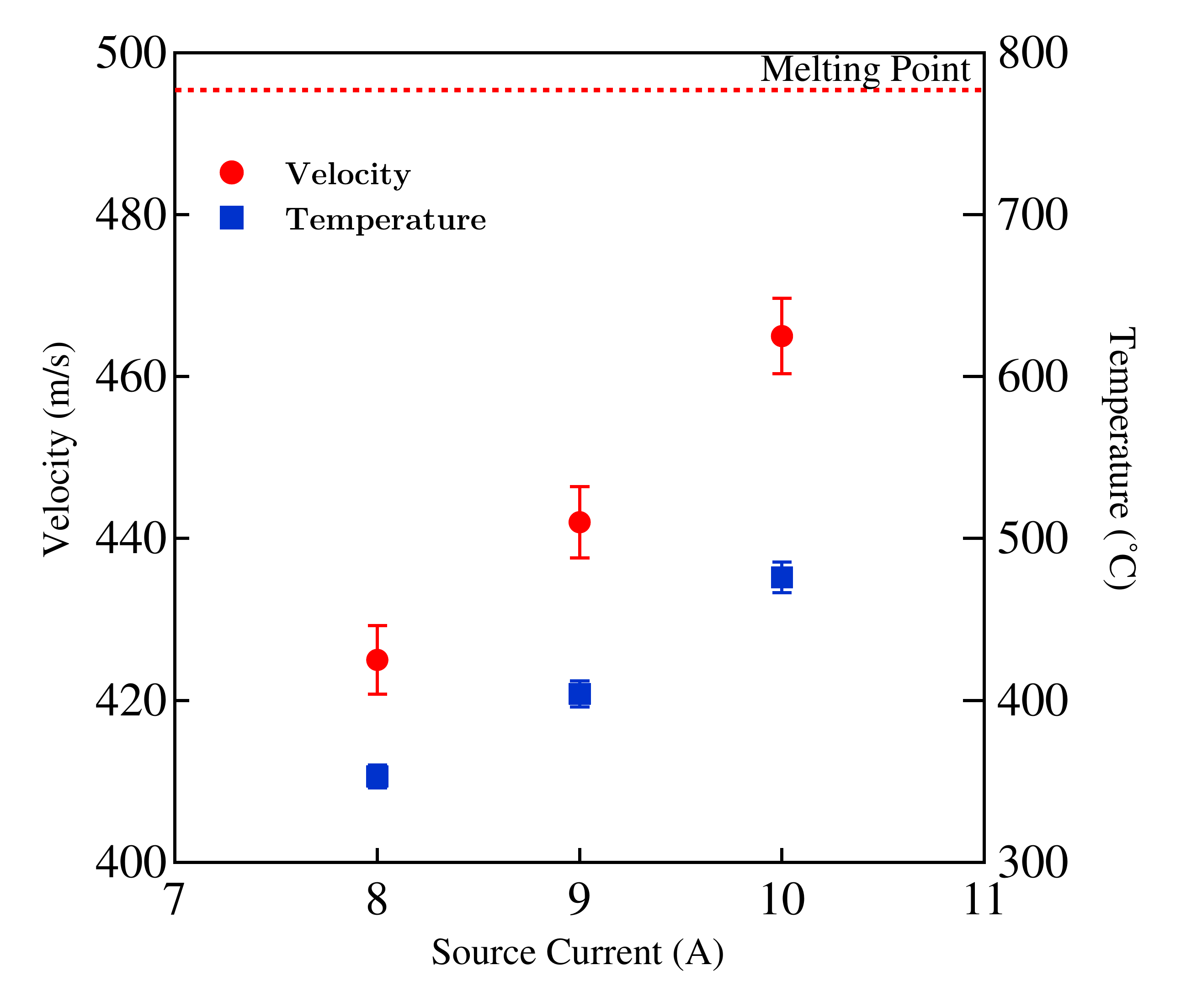}  
	
	\caption{Most probable velocity of atoms  (red filled circles) and the estimated temperature (filled blue squares) of the dispenser vs. the current through the dispenser:  A laser beam crossing the atomic beam at an angle  4.7${^\circ}$ is used to probe the velocity distribution of the atomic beam. The broken horizontal line indicates the melting point of Sr.}
\label{fig:d2} 
\end{figure}

The fluorescence is collected through a rectangular slit  by a plano convex lens of focal length 75 mm on a photomultiplier tube (Hamamatsu Model H9307-02) in the perpendicular direction to probe beam. We scan the probe beam frequency over a range of 700 MHz by applying a linear voltage ramp to the piezoelectric transducer of the seed laser (922 nm) to cover the spectrum of all isotopes of Sr.
The obtained fluorescence signal is fitted with a sum of six Voigt profiles for three Bosonic and three hyperfine components of Fermionic $^{87}$Sr isotope as shown in Fig. 3. The relative area under each peak of the fluorescence signal is proportional to the relative abundance of isotopes.\par
By using the known values of the isotope shifts, hyperfine splitting, relative abundances \cite{lineshifts} and considering a power and pressure broadened Lorentzian width ($\Gamma$ = 42 MHz), we perform a multiple Voigt fitting to the measured signal (similar to Ref. \cite{bridge2009vapor}). We estimate the transverse Doppler line broadening of each component, offset and center. We use the relative peak positions of the various spectral lines for frequency scaling and  alternately, we also cross check the frequency calibration against the known Free Spectral Range (FSR) of the frequency doubling optical cavity.
\begin{figure}[hbt]
\centering
\includegraphics[width=0.75\linewidth]{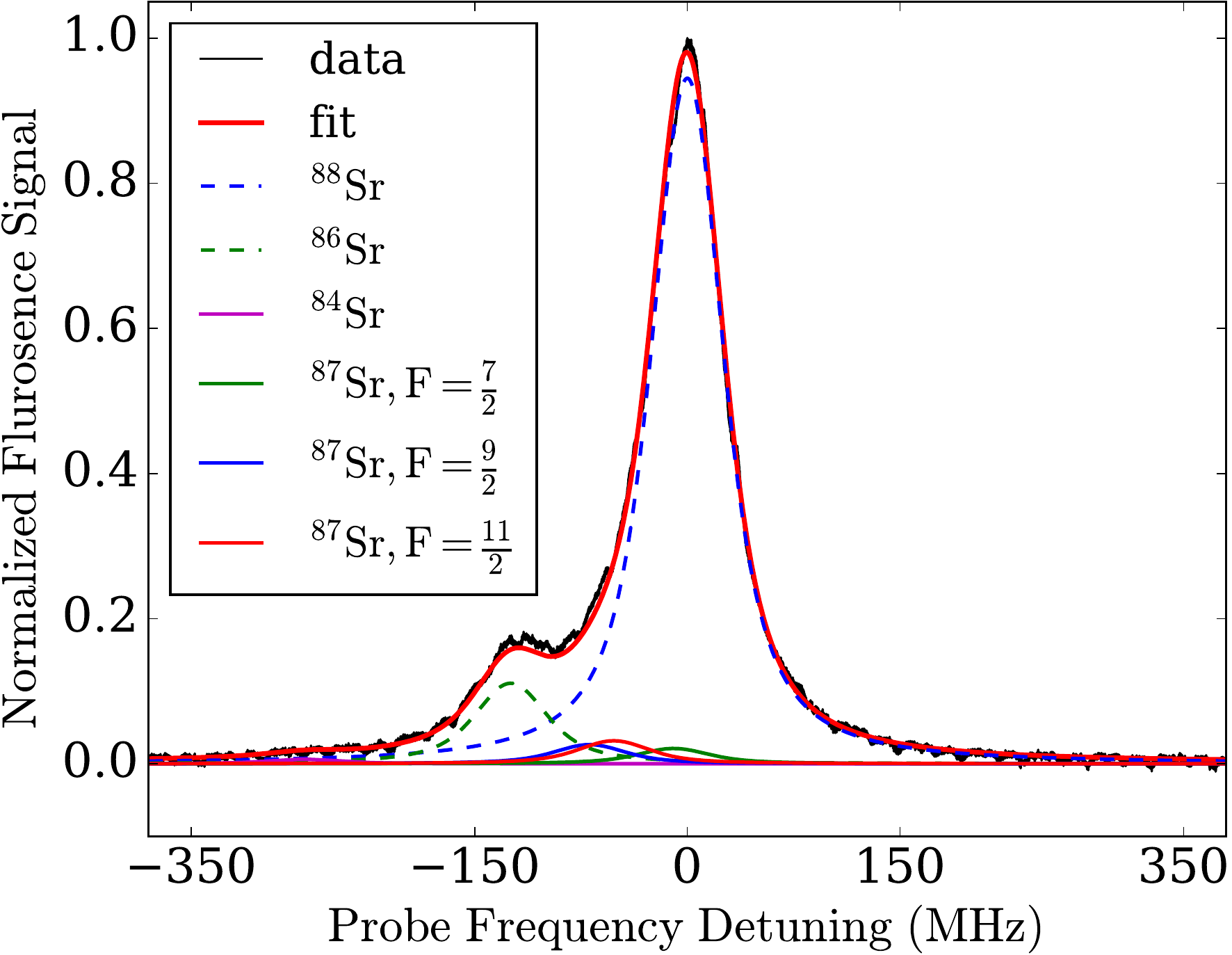}  
	
	\caption{Fluorescence spectrum of $5s^2\ ^1S_0\longrightarrow 5s\ 5p\ ^1P_1$ transition of Sr atoms. The spectrum is fitted with a sum of six Voigt profiles for the individual isotopes of Sr.}
\label{fig:v} 
\end{figure}

The estimated width of the transverse velocity profile is determined by fitting  the florescence peaks recorded for a dispenser source current of 8A. The estimated $^{88}$Sr  full-Width-half-maximum (FWHM) and corresponding velocity spread is found to be  $33.3 \pm 1.7$ MHz, $6.5 \pm 0.3$ m/s respectively which is reasonably less compared to reported results in previous studies. The transverse Doppler width is found to be  very close to natural linewidth ($\Gamma$ = 32 MHz) which is essentially equivalent to a Doppler free spectrum. \par
Continuous operation of Strontium source for 6-7 hour at higher currents ($>$ 9 A) during the experiment  increases the background pressure in the spectroscopy cell which broadens the spectral lines. To maintain high vacuum pressure in the cell, we implemented a TSP as shown in Fig.1. We tested the performance of the TSP in maintaining the pressure by monitoring the FWHM of $^{88}$Sr peak over time while keeping the dispenser source continuously ON for 5-6 hours on three consecutive days. We operated the source at 10 A current and kept it ON continuously for over 6 hours. We observe that the FWHM  continuously increase  from 34.2 MHz for the initial 3 hrs and then saturates at 54.8 MHz, while the overall flux remains almost constant. This velocity broadening can be attributed to an increased background pressure. 
After 6 hours we turned OFF the source and switched ON the TSP for half an hour by passing 5A current through it. Next day we repeated the measurement to observe the similar behavior of FWHM, we saw that the FWHM dropped to the initial value (35.6 MHZ) which indicates the regain of vacuum inside the cell also a quick saturation of FWHM to a lower value (44.2 MHz) over 5.3 hours run compared to previous day. Similar behavior was seen on third day, where we see FWHM going from 35.2 MHz to 36.5 MHz. This exercise not only indicate the reasonably good performance of TSP but also an active pumping of TSP during the continuous operation of source. We have been using this source continuously for over 3 months without any degradation in the observed spectral qualities.\par  

In conclusion, we have demonstrated a simple, compact, light weight, low power solution for performing  spectroscopy of Sr atoms. The use of beam spectroscopy, reduces the optical complexity of SAS and gives an equally well resolved Doppler free spectrum of Sr which can be used for stabilization of Laser frequency and other spectroscopic experiments. The construction of this system only uses standard glass blowing techniques in combination with a compact TSP.\par   

The authors would like to thank the Department of Science and Technology, Govt. of India for grants through  from EMR/2014/000365 and Nano Mission. The authors also wish to thank the help of Mr. Nair with glass blowing. 

  \bibliographystyle{apsrev}
\bibliography{Ref.bib}
\end{document}